\documentclass{sf2a-conf2015}
\usepackage{txfonts}
\usepackage{graphicx}
\usepackage{hyperref}
\usepackage[]{natbib}  
\usepackage{epstopdf}

\def\BibTeX{{\rm B\kern-.05em{\sc i\kern-.025em b}\kern-.08em
    T\kern-.1667em\lower.7ex\hbox{E}\kern-.125emX}}
\bibpunct{(}{)}{;}{a}{}{,}  

\begin{document}

\TitreGlobal{SF2A 2015}


\title{Melting the core of giant planets: impact on tidal dissipation}

\runningtitle{Melting the core of giant planets: impact on tidal dissipation}

\author{S. Mathis}\address{Laboratoire AIM Paris-Saclay, CEA/DSM - CNRS - Universit\'e Paris Diderot, IRFU/SAp Centre de Saclay, F-91191 Gif-sur-Yvette Cedex, France}

\setcounter{page}{237}


\maketitle


\begin{abstract}
Giant planets are believed to host central dense rocky/icy cores that are key actors in the core-accretion scenario for their formation. In the same time, some of their components are unstable in the temperature and pressure regimes of central regions of giant planets and only ab-initio EOS computations can address the question of the state of matter. In this framework, several works demonstrated that erosion and redistribution of core materials in the envelope must be taken into account. These complex mechanisms thus may deeply modify giant planet interiors for which signatures of strong tidal dissipation have been obtained for Jupiter and Saturn. The best candidates to explain this dissipation are the viscoelastic dissipation in the central dense core and turbulent friction acting on tidal inertial waves in their fluid convective envelope. In this work, we study the consequences of the possible melting of central regions for the efficiency of each of these mechanisms.
\end{abstract}



\section{Introduction and context}

While more and more giant gaseous and icy planets are discovered in exoplanetary systems \citep[e.g.][]{Perryman2011}, internal structure of Jupiter, Saturn, Uranus and Neptune are still uncertain \citep[e.g.][]{Guillot1999,FBM2011,Baraffeetal2014}. More particularly, the mass and the radius of potential central dense cores of rocks and ices are still unknown \citep[e.g.][]{MH2009,Nettelmann2011,HelledGuillot2013} with no firm constrains \citep[][]{Hubbardetal2009,Gaulmeetal2011} while they are key elements for core-accretion formation scenario of these planets \citep[e.g.][]{Pollacketal1996,Morbietal2015}. In this framework, several works have demonstrated that elements constituting these regions (as for example silicates like MgSiO$_3$) are thermodynamically unstable in the pressure and temperature regimes of giant planet interiors \citep[e.g.][]{WilsonMilitzer2012,WilsonMilitzer2012b,Wahletal2013,Getal2014,Mazevetetal2015}. These results demonstrate that central dense regions constituted of heavy elements may dissolve in the envelope. It may lead to an erosion of the core, a redistribution of core materials in the envelope and a modification of the nature of the core-envelope boundary that impact the structure and evolution of giant planets.\\

At the same time, tidal interactions are one of the key mechanisms that drive the evolution of planetary systems. On one hand, in the solar system, \cite{Laineyetal2009} and \cite{Laineyetal2012} have provided new constrains on tidal dissipation in Jupiter and Saturn using high precision astrometry. The obtained values are stronger than those that have been previously proposed in the literature by one order of magnitude, i.e. $Q'_{\rm J}\equiv10^{5.15}$ and $Q'_{\rm S}\equiv10^{3.87}$, where $Q'$ is the normalized tidal quality factor defined in \cite{OL2007}, which is inversely proportional to tidal dissipation. On the other hand, constrains obtained in exoplanetary systems hosting Hot Jupiters lead to a weaker dissipation with $Q'_{\rm HJ}\equiv10^{6.5}$ \citep[see][and references therein]{Ogilvie2014}. In giant planets, physical mechanisms driving tidal dissipation are the viscoelastic dissipation in the potential dense central region constituted of rocks and ices \citep[][]{Remusetal2012,Remusetal2015} and the turbulent friction acting on tidal inertial waves (their restoring force is the Coriolis acceleration) in the deep convective fluid envelope \citep[][]{OgilvieLin2004}. The efficiency of both mechanisms depends on the mass and radius aspect ratios between the core and the envelope and on the nature of the core-envelope boundary \citep[][]{GoodmanLackner2009,Ogilvie2013,GMR2014}. It is thus necessary to examine the possible consequences of the melting of the core. First, in Sec. \ref{sec:2}, we discuss the direct impact of the modification of the mass and the size of the core for each dissipation processes. Next, in Sec. \ref{sec:3}, we study the impact of the core-envelope boundary for tidal inertial waves propagating in the fluid envelope. Finally, we conclude and discuss astrophysical consequences and perspectives of this work.

\section{Tidal dissipation mechanisms and their dependences on the mass and the size of the core}
\label{sec:2}

Let us first consider simplest bi-layer models of giant planets constituted by an external fluid envelope and a solid core constituted of rocks and ices (see Fig. \ref{Structure}, left-panel). In such a simplified model, two dissipation mechanisms for tidal friction have been identified and studied:

\begin{figure}[t!]
\begin{center}
\includegraphics[width=0.65\linewidth]{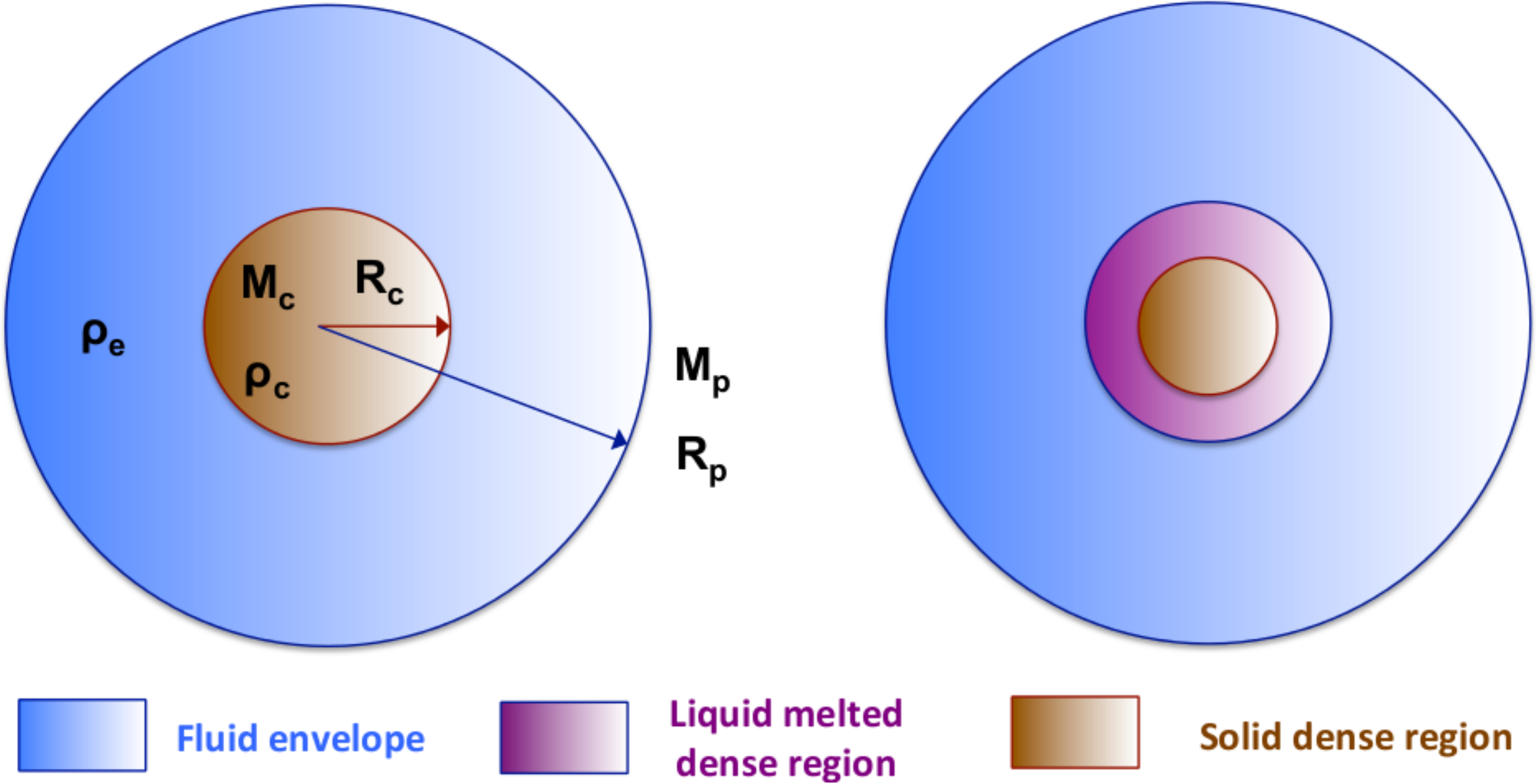}
\end{center}
\caption{{\bf Left:} simplest bi-layer planetary model with a central solid (rocky/icy) core; $R_{\rm c}$ ($R_{\rm p}$) and $M_{\rm c}$ ($M_{\rm p}$) are the radius and the mass of the core (the planet) while $\rho_{\rm e}$ and ${\rho_{\rm c}}$ are the density of the envelope and of the core respectively. {\bf Right:} Three layer planetary model inspired by the results obtained by \cite{Mazevetetal2015} in the case where central regions melting is occurring.}
\label{Structure}
\end{figure}

\begin{itemize} 
\item {\bf The inelastic dissipation:} Rocks and ices can be modeled as viscoelastic materials. Because of the uncertainties on their rheological behavior, the simplest assumption is to assume the linear Maxwell model for which the shear modulus is complex with
\begin{equation}
\label{eq:maxwell}
{\rm Re}\left[\bar{\mu}(\omega)\right] = \frac{\eta^2 \, G \, \omega^2}{G^2+\eta^2 \, \omega^2}\quad\hbox{and}\quad {\rm Im}\left[\bar{\mu}(\omega)\right]= \frac{\eta \, G^2 \, \omega}{G^2+\eta^2 \, \omega^2} ,
\end{equation}
where $\omega$ is the tidal frequency, $G$ the rigidity and $\eta$ the viscosity that drives the friction \citep{Remusetal2012,Remusetal2015}. As demonstrated in \cite{Ogilvie2013} and \cite{GMR2014}, it is possible to evaluate a frequency-averaged tidal dissipation, which allows to unravel the impact of internal structure on tidal dissipation. In the case of a viscoelastic core, we obtain: 
\begin{equation}
\left<{\mathcal D_{\rm ve}}\right>_{\omega}=\int^{+\infty}_{-\infty} \! {\rm Im} \left[k_2^2(\omega)\right] \,\frac{\mathrm{d}\omega}{\omega}  = \frac{\pi \,G \left(3 + 2{\mathcal A}\right)^2{\mathcal B}\,{\mathcal C}}{{\mathcal D}\left(6\,{\mathcal D}+4\,{\mathcal A}{\mathcal B}{\mathcal C}G\right)},
\label{inelastic}
\end{equation}
with
\begin{equation}
{\mathcal A}=1 + \frac{5}{2}\gamma^{-1}\alpha^{3}\left(1-\gamma\right),\quad{\mathcal B} = \alpha^{-5}\left(1-\gamma\right)^{-2}, \quad {\mathcal C}=\frac{19}{2 \, \rho_{\rm c} \, g_{\rm c} \, R_{\rm c}},\quad{\mathcal D}= \left[\frac{2}{3}\,{\mathcal A}{\mathcal B}\left(1 - \gamma\right) \left(1 + \frac{3}{2} \gamma\right)-\frac{3}{2}\right].
\end{equation}
We introduce
\begin{equation}
\alpha=\frac{R_{\rm c}}{R_{\rm p}}\hbox{,}\quad\beta=\frac{M_{\rm c}}{M_{\rm p}}\quad\hbox{and}\quad\gamma=\frac{\rho_{\rm e}}{\rho_{\rm c}}=\frac{\alpha^3\left(1-\beta\right)}{\beta\left(1-\alpha^3\right)}<1,
\end{equation}
where $R_{\rm c}$ ($R_{\rm p}$) and $M_{\rm c}$ ($M_{\rm p}$) are the radius and the mass of the core (the planet) while $\rho_{\rm e}$ and ${\rho_{\rm c}}$ are the density of the envelope and of the core respectively. We introduce the quadrupolar complex Love number $k_{2}^{2}$, associated with the $\left(2,2\right)$ component of the time-dependent tidal potential $U$ that corresponds to the spherical harmonic $Y_2^2$. It quantifies at the surface of the planet ($r=R_{\rm p}$) the ratio of the tidal perturbation of its self-gravity potential over the tidal potential in the simplest case of coplanar systems. In the case of dissipative systems, it is a complex quantity which depends on the tidal frequency ($\omega$) with a real part that accounts for the energy stored in the tidal perturbation, while the imaginary part accounts for the energy losses \citep[e.g.][]{Remusetal2012}. 

\item {\bf The (viscous) turbulent friction applied on tidal inertial waves:} if we assume that the external fluid envelope is convective, the presence of a close natural satellite or star excites tidal inertial waves if $\omega\in\left[-2\Omega,2\Omega\right]$, where $\Omega$ is the angular velocity of the planet. Their restoring force is the Coriolis acceleration. Because of the friction applied by convective turbulence, their kinetic energy is converted into heat that leads to tidal evolution of planet-star/natural satellite systems. \cite{Ogilvie2013} derived the corresponding frequency-averaged dissipation
\begin{equation}
\left<{\mathcal D}_{\rm in}^{\rm s}\right>_{\omega}=\int^{+\infty}_{-\infty} \! {\rm Im} \left[k_2^2(\omega)\right] \,\frac{\mathrm{d}\omega}{\omega} = \frac{100 \pi}{63} \epsilon^2 \displaystyle{\frac{\alpha^5}{1-\alpha^5}}
\left[ 1+ \frac{1-\gamma}{\gamma}\alpha^3 \right]\left[ 1+ \frac{5}{2} \frac{1-\gamma}{\gamma}\alpha^3 \right]^{-2}
\label{fluid_solid}
\end{equation}
\end{itemize}
assuming a solid dense core on which tidal waves reflect. It may lead to inertial wave attractors where viscous friction is efficient \citep[][]{O2005}. We introduce $\epsilon^2 \equiv \left(\Omega/ \sqrt{\mathcal{G} M_{\rm p} / R_{\rm p}^3}\right)^2=\left(\Omega/\Omega_{\rm c}\right)^2 \ll 1$, where $\Omega_{\rm c}$ {is the critical angular velocity} and ${\mathcal G}$ is the gravitational constant. We also define the normalized frequency-averaged tidal dissipation at fixed angular velocity
\begin{equation}
\left<{\mathcal D}_{\rm in}^{\rm s}\right>_{\omega}^{\Omega}=\epsilon^{-2}\left<{\mathcal D}_{\rm in}^{\rm s}\right>_{\omega}.
\end{equation}\\

\begin{figure}[!t]
\begin{center}
\includegraphics[width=0.4\linewidth]{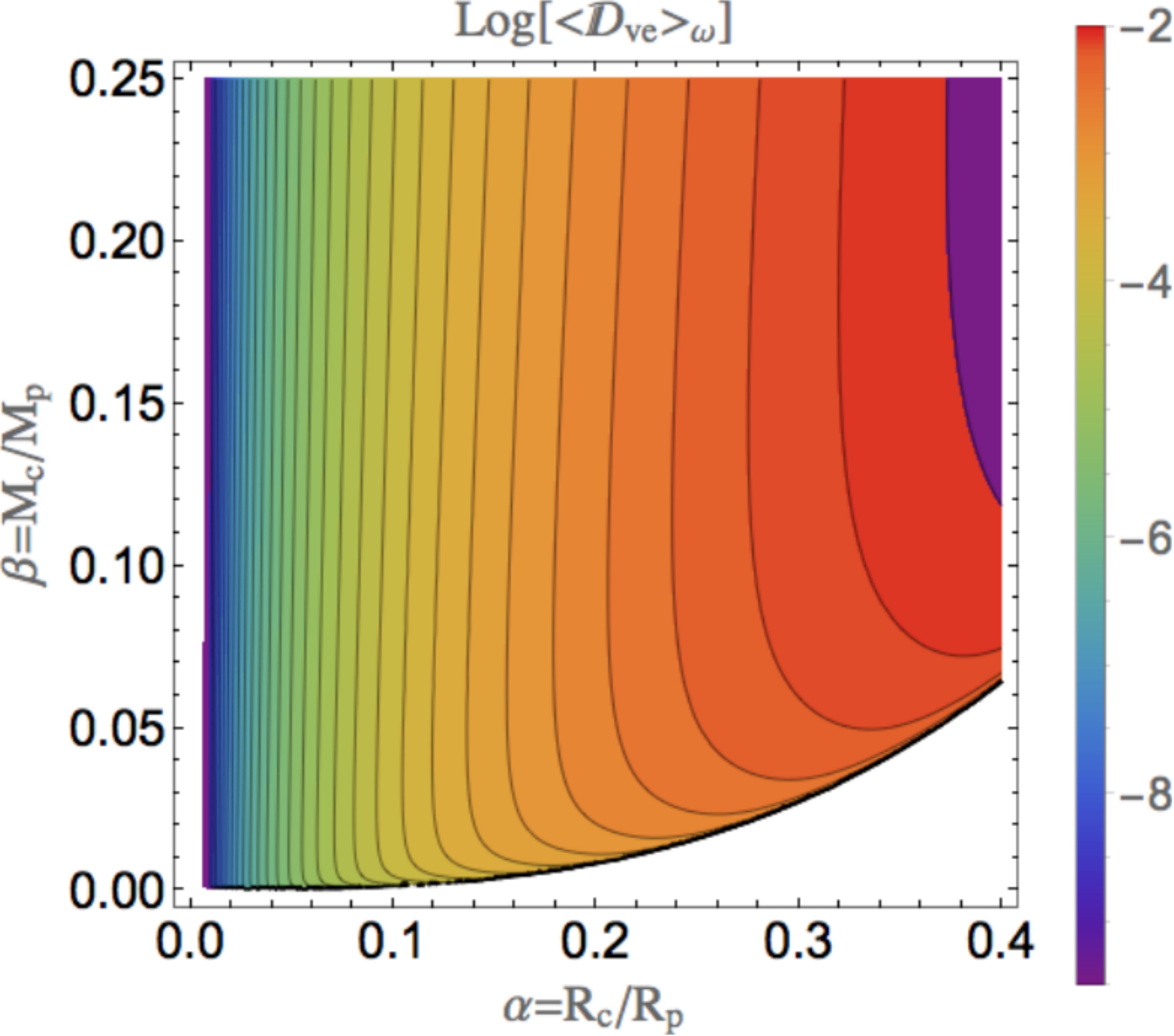}\quad\quad\quad
\includegraphics[width=0.4\linewidth]{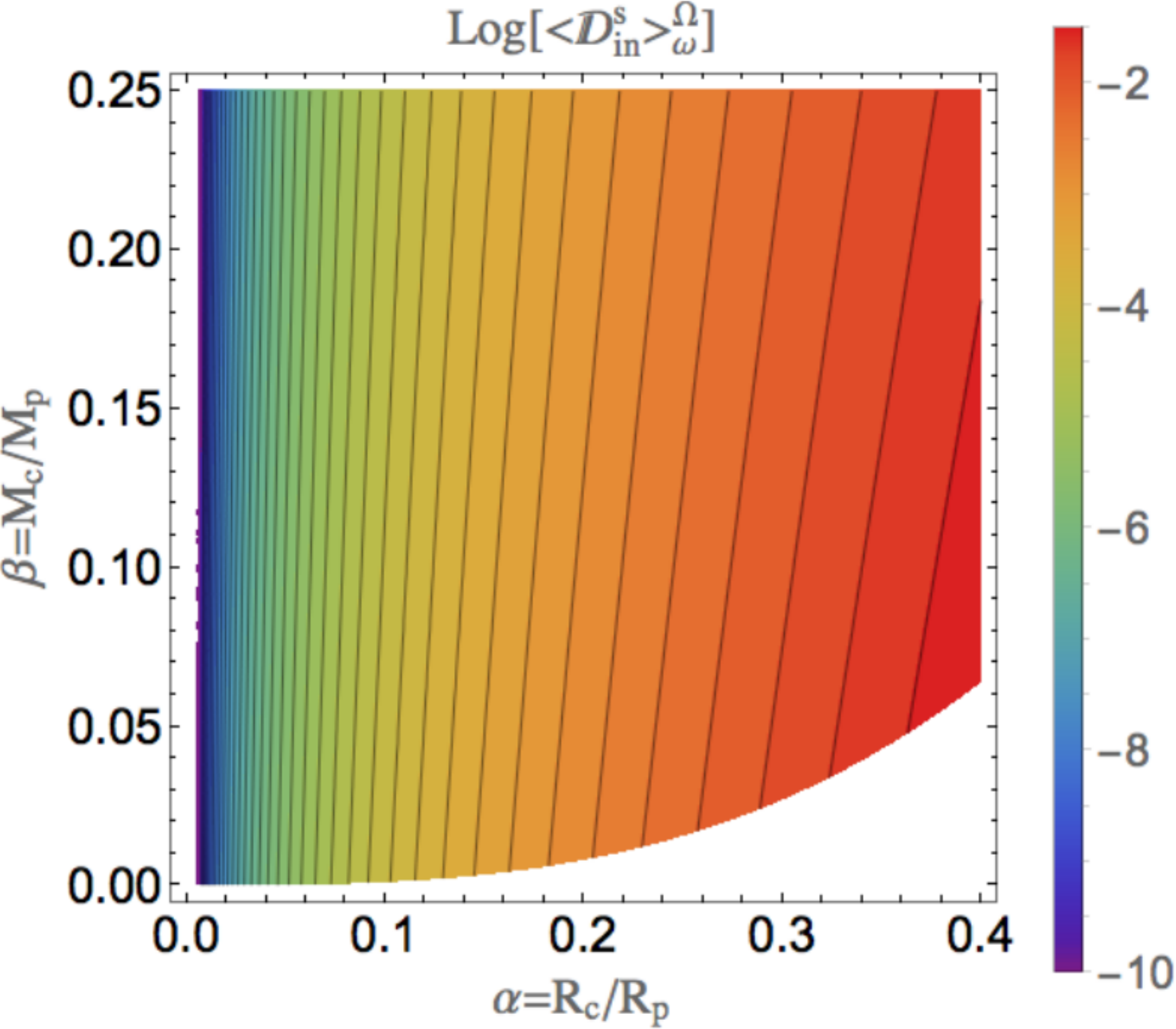}
\includegraphics[width=0.4\linewidth]{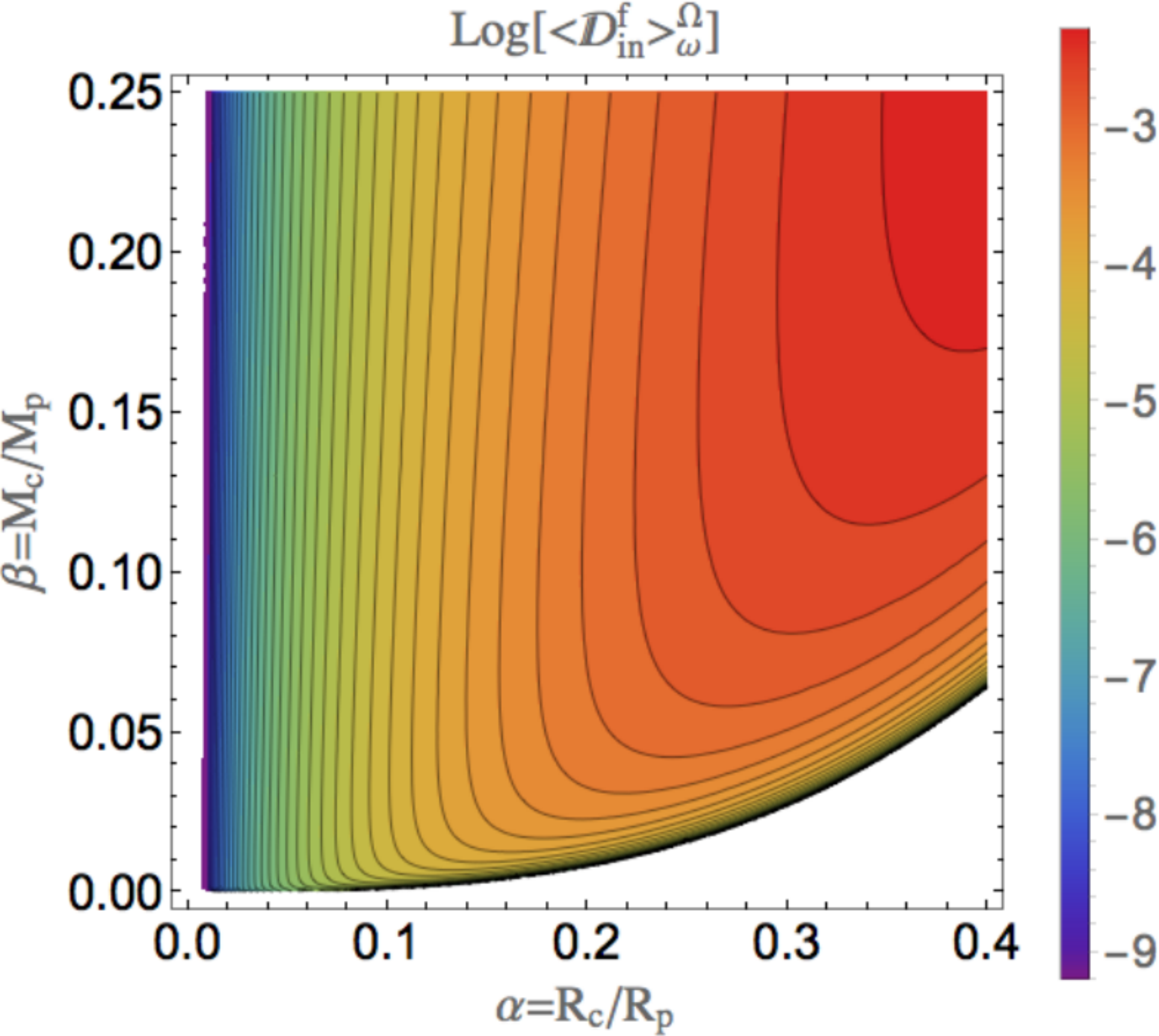}\quad\quad\quad
\includegraphics[width=0.4\linewidth]{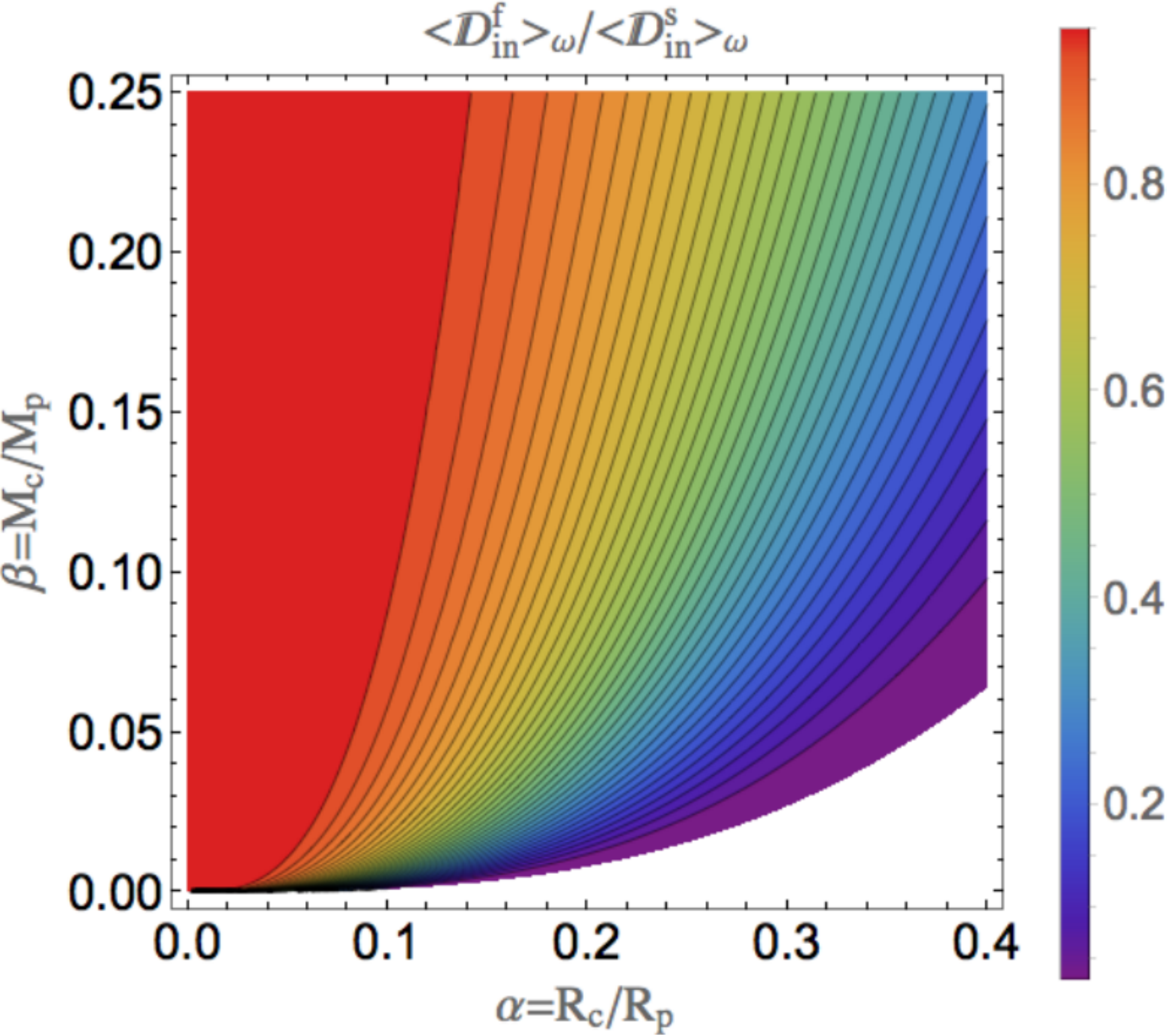}
\end{center}
\caption{Variations of $\left<{\mathcal D_{\rm ve}}\right>_{\omega}$ (left-top panel), $\left<{\mathcal D}_{\rm in}^{\rm s}\right>_{\omega}^{\Omega}$ (right-top panel), $\left<{\mathcal D}_{\rm in}^{\rm f}\right>_{\omega}^{\Omega}$ (left-bottom panel) and $\left<{\mathcal D}_{\rm in}^{\rm f}\right>_{\omega}/\left<{\mathcal D}_{\rm in}^{\rm s}\right>_{\omega}$ (right-bottom panel) as a function of the radius and mass aspect ratios ($\alpha=R_{\rm c}/R_{\rm p}$ and $\beta=M_{\rm c}/M_{\rm p}$ respectively).}
\label{Maps}
\end{figure}

{\bf Looking for tidal evolutionary track:} In Fig. \ref{Maps}, we represent the variation of $\left<{\mathcal D_{\rm ve}}\right>_{\omega}$ (assuming the parameters used in \cite{GMR2014} for Saturn-like planets) and $\left<{\mathcal D}_{\rm in}^{\rm s}\right>_{\omega}^{\Omega}$ as a function of the radius and mass ratios $\alpha=R_{\rm c}/R_{\rm p}$ and $\beta=M_{\rm c}/M_{\rm p}$ (respectively in top-left and top-right panels). First, the mass of the material in solid state and thus $\beta$ is decreased when melting occurs. Next,  because of macroscopic transport of heavy elements towards the envelope, its mean density ($\rho_{\rm e}$) is increased while the one of the core ($\rho_{\rm c}$) is decreased (at a fixed core mass, it corresponds to an increase of $R_{\rm c}$). To be able to predict the effect of melting processes, it would thus be necessary to be able to draw planetary evolutionary track in the $\left(\alpha,\beta\right)$ plane and in the corresponding $\left(\beta,\gamma\right)$ one. Such a methodology has been applied in \cite{Mathis2015} in the case of rotating low-mass stars leading to robust results for the frequency-averaged tidal dissipation in their convective envelope. To compute such a {\it tidal evolutionary track}, it would be necessary to compute planetary structure and evolution models taking into account melting and related mixing mechanisms \citep[e.g.][]{LC2012} and the most realistic available EOS \citep[e.g.][]{Mazevetetal2015}. Note also that the value of $G$ depends on the state of materials.

\section{The impact of the core-envelope boundary condition}
\label{sec:3}

In addition of modifying the radius and mass aspect ratios between the central dense core and the envelope, melting the core of gaseous and icy giant planets may also modify the nature of the boundary condition between these two regions. Indeed, let us focus here on tidal dissipation in the deep hydrogen and helium envelope in two configurations: i) in the first, we consider it surrounds a dense {\it solid} rocky/icy core (see Fig. \ref{Structure}, left panel); ii) in the second one, we consider the configuration suggested by \citep{Mazevetetal2015} where the envelope is above a {\it liquid} shell of SiO$_2$ that surrounds a central core of MgO (see Fig. \ref{Structure}, right panel).\\ 

Then, tidal dissipation through the viscous dissipation of inertial waves is not the same in the case of a {\it fluid-solid interface} and of a {\it fluid-fluid interface}. As in the previous paragraph, we consider the behavior of the frequency-averaged tidal dissipation as introduced by \cite{Ogilvie2013} and computed in the case of gaseous giant planets by \cite{GMR2014}. 

In the case of a fluid-solid interface, the frequency-averaged tidal dissipation is given by  Eq. (\ref{fluid_solid}) while in the case of a fluid-fluid interface it becomes
\begin{eqnarray}
\lefteqn{\left<{\mathcal D}_{\rm in}^{\rm f}\right>_{\omega}=\int^{+\infty}_{-\infty} \! {\rm Im} \left[k_2^2(\omega)\right] \,\frac{\mathrm{d}\omega}{\omega} = \frac{100 \pi}{63} \epsilon^2 \left(\frac{\alpha^5}{1-\alpha^5}\right)\left(1-\gamma\right)^2}\nonumber\\
&&\times\left(1-\alpha\right)^4\left(1+2\alpha+3\alpha^2+\frac{3}{2}\alpha^3\right)^2\left[1+\left(\frac{1-\gamma}{\gamma}\right)\alpha^3\right]\left[1+\frac{3}{2}\gamma+\frac{5}{2\gamma}\left(1+\frac{1}{2}\gamma-\frac{3}{2}\gamma^2\right)\alpha^3-\frac{9}{4}\left(1-\delta\right)\alpha^5\right]^{-2}.
\label{fluid_fluid}
\end{eqnarray}
We also introduce the normalized frequency-averaged tidal dissipation at fixed angular velocity
\begin{equation}
\left<{\mathcal D}_{\rm in}^{\rm f}\right>_{\omega}^{\Omega}=\epsilon^{-2}\left<{\mathcal D}_{\rm in}^{\rm f}\right>_{\omega}.
\end{equation}\\

In Fig. \ref{Maps} (left-bottom panel), we represent its variation as a function of $\alpha=R_{\rm c}/R_{\rm p}$ and $\beta=M_{\rm c}/M_{\rm p}$. We immediately see that its behavior is different that the one of $\left<{\mathcal D}_{\rm in}^{\rm s}\right>_{\omega}^{\Omega}$. First, as observed in \cite{Mathis2015}, it is maximum around $\alpha\approx0.571$ and $\beta\approx0.501$. Moreover, the maximum of its amplitude is weaker than those of $\left<{\mathcal D}_{\rm in}^{\rm s}\right>_{\omega}^{\Omega}$. For these reason, we plot in Fig. \ref{Maps} (right-bottom panel) the ratio $\left<{\mathcal D}_{\rm in}^{\rm f}\right>_{\omega}^{\Omega}/\left<{\mathcal D}_{\rm in}^{\rm s}\right>_{\omega}^{\Omega}=\left<{\mathcal D}_{\rm in}^{\rm f}\right>_{\omega}/\left<{\mathcal D}_{\rm in}^{\rm s}\right>_{\omega}$ as a function of $\alpha$ and $\beta$. Having a fluid-fluid interface then decreases the strength of tidal dissipation in the external envelope. It is then interesting to evaluate $\left<{\mathcal D}_{\rm in}^{\rm f}\right>_{\omega}/\left<{\mathcal D}_{\rm in}^{\rm s}\right>_{\omega}$ for possible values of $M_{\rm c}$ and $R_{\rm c}$ for Jupiter-, Saturn-, Uranus-, and Neptune-like planets; the obtained results and the used parameters are given in the following table \citep{Guillot1999,Hubbardetal2009,Helledetal2011,PH2012,Nettelmannetal2013}.\\

\begin{table}[h!]
  \centering
  \begin{tabular}[center]{cccccc}\hline \hline
    Parameter & Jupiter & Saturn & Uranus & Neptune \\ \hline
    \\
    	$M_{\rm p}$ ($M_{\rm E}$) & $317.830$ & $95.159$  & $14.536$  & $17.147$ \\
    \\
        $R_{\rm p}$ ($R_{\rm E}$) & $10.973$ & $9.140$  & $3.981$  & $3.865$ \\
    \\
        $M_{\rm c}$ ($M_{\rm E}$)  & $6.41$ & $18.65$  & $1.35$  & $2.25$ \\
    \\
        $\alpha=R_{\rm c}/R_{\rm p}$  & $0.126$ & $0.219$  & $0.30$  & $0.35$ \\
    \\
        $\beta=M_{\rm c}/M_{\rm p}$  & $0.020$ & $0.196$  & $0.093$  & $0.131$ \\
    \\    
    \hline\\
         $\left<{\mathcal D}_{\rm in}^{\rm f}\right>_{\omega}/\left<{\mathcal D}_{\rm in}^{\rm s}\right>_{\omega}$ & $0.612$ & $0.766$  & $0.237$  & $0.194$ \\
    \\
    \hline\\
\end{tabular}
  \caption{Values of the parameters used to compute the ratio $\left<{\mathcal D}_{\rm in}^{\rm f}\right>_{\omega}/\left<{\mathcal D}_{\rm in}^{\rm s}\right>_{\omega}$ \citep[][]{Remusetal2012,Helledetal2011,PH2012,Nettelmannetal2013}}
  \label{tab:fluid_solid}
\end{table}
Two conclusions are then obtained. First, for every planets the frequency-averaged tidal dissipation associated with inertial waves reflecting on a fluid-fluid interface is weaker than the one in the case of a fluid-solid interface. Next, as this is shown by Fig. \ref{Maps} (right-bottom panel), this decrease is much stronger in the case of planets with large radius aspect ratios, that corresponds to the case of icy giant planets for which the strength of the frequency-averaged tidal dissipation corresponds to only $20\%$ of its value in the case of a fluid-solid interface.

\section{Conclusions}

In this first exploratory work, we have discussed the possible effects of melting mechanisms that may occur in the central dense regions of giant planets. First, we have shown that realistic planetary structure and evolution models are key ingredients to be developed to provide a precise evaluation of mass, radius, and density ratios necessary to get a robust prediction of tidal dissipation. Indeed, these quantities directly impact the amplitude of the frequency-averaged tidal dissipation. Moreover, if melting of rocks and ices is taking place, tidal inertial waves, which propagate in the external envelope, may reflect on a {\it fluid-fluid} interface instead of a {\it fluid-solid} interface. In the case of planets having a large core such as icy giant planets, this may lead to a net decrease of tidal dissipation. In a near future, the impact of mixing mechanisms such as double-diffusive instabilities \citep[see e.g][]{LC2012} on tidal dissipation must also be evaluated.

\begin{acknowledgements}
The author acknowledges funding by the European Research Council through ERC grant SPIRE 647383. This work was also supported by the Programme National de Plan\'etologie (CNRS/INSU) and the International Space Institute (ISSI team ENCELADE 2.0).
\end{acknowledgements}

\bibliographystyle{aa}  
\bibliography{Mathis2} 

\end{document}